\documentclass[aps,prb, amsfonts, amssymb, amsmath, superscriptaddress, notitlepage,twocolumn,10pt,nobalancelastpage]{revtex4-2}
\usepackage{graphicx} 
\usepackage{hyperref}
\usepackage{bm}
\usepackage{textgreek}
\usepackage{soul}
\hypersetup{ 
    colorlinks=true, %set true if you want colored links
    linktoc=all,     %set to all if you want both sections and subsections linked
    linkcolor=blue,  %choose some color if you want links to stand out
    citecolor =blue
} 
\usepackage{amsmath}

\newcommand{\bk}{\mathbf{k}}
\newcommand{\bq}{\mathbf{q}}
\newcommand{\br}{\mathbf{r}}
\newcommand{\bd}{\bm{\delta}}

\newcommand{\bK}{\mathbf{K}}
\newcommand{\bQ}{\mathbf{Q}}

\newcommand{\bM}{\mathbf{M}}
\newcommand{\bG}{\mathbf{G}}

\newcommand{\bn}{\mathbf{n}}
\newcommand{\ba}{\mathbf{a}}

\newcommand{\ez}{\hat{\mathbf{e}}_z}

\newcommand{\nn}{\nonumber}

\begin{document}
\title{Nature of the Topological Transition of the Kitaev Model in [111] Magnetic Field}

\author{S. Thiagarajan}
\affiliation{London Centre for Nanotechnology, University College London, Gordon St., London, WC1H 0AH, United Kingdom}
\author{C. Watson}
\affiliation{London Centre for Nanotechnology, University College London, Gordon St., London, WC1H 0AH, United Kingdom}
\author{T. Yzeiri}
\affiliation{London Centre for Nanotechnology, University College London, Gordon St., London, WC1H 0AH, United Kingdom}
\affiliation{ISIS Facility, Rutherford Appleton Laboratory, Chilton, Didcot, Oxfordshire OX11 0QX, United Kingdom}
\author{H. Hu}
\affiliation{School of Physics and Astronomy, University of Birmingham, Edgbaston Park Road, Birmingham, B15 2TT, United Kingdom}
\author{B. Uchoa}
\affiliation{Department of Physics and Astronomy, University of Oklahoma, Norman, Oklahoma 73069, USA}
\author{F. Kr\"uger}
\affiliation{London Centre for Nanotechnology, University College London, Gordon St., London, WC1H 0AH, United Kingdom}
\affiliation{ISIS Facility, Rutherford Appleton Laboratory, Chilton, Didcot, Oxfordshire OX11 0QX, United Kingdom}

\begin{abstract}
We investigate the nature of the topological phase transition of the antiferromagnetic Kitaev model on the honeycomb lattice in 
the presence of a magnetic field along the [111] direction. The field opens a topological gap in the Majorana fermion spectrum and leads to a sequence 
of topological phase transitions before the field polarised state is reached. At mean field level the gap first closes at the three $M$ points in the Brillouin zone, 
where the Majorana fermions form Dirac cones, resulting in a change of Chern number by three.   An odd number of Dirac fermions in the infrared is unusual 
and requires Berry curvature compensation in the UV, which occurs via topological, ring-like hybridisation gaps with higher-energy bands.
We perform a renormalisation-group analysis of the topological phase transition 
at the three $M$ points within the Yukawa theory, allowing for intra- and inter-valley fluctuations of the spin-liquid bond operators. We find that the latter  
lead to a breaking of Lorentz invariance and hence a different universality compared to the standard Ising Gross-Neveu-Yukawa class.  
\end{abstract}
 
\maketitle

\section{Introduction}
\label{sec.intro}
The celebrated Kitaev honeycomb model, a bond-dependent Ising model, has an exactly solvable quantum spin liquid (QSL) ground state after the 
spin-1/2 operators are fractionalised into Majorana fermions \cite{Kitaev+2006}. Many efforts to realise this model in materials have utilised spin–orbit 
coupling, as proposed by Jackeli and Khaliullin \cite{Jackeli+2009}, for which honeycomb iridates \cite{Takayama+2015,Choi+2012,Chun+2015} and 
\textalpha-RuCl\textsubscript{3} \cite{Warzanowski+2020,Plumb+2014,Zhou+2023} are promising candidates. However, these materials display long-range 
zigzag antiferromagnetic (AFM) order at low temperatures \cite{Suzuki+2021,Liu+2011,Ye+2012}, which can be suppressed in \textalpha-RuCl\textsubscript{3} 
by applying a magnetic field \cite{Sears+2017,Wolter+2017}.

These observations have motivated various theoretical and numerical investigations of the AFM Kitaev honeycomb model with an applied field.
As already demonstrated in the seminal work by Kitaev \cite{Kitaev+2006}, in third order perturbation theory a small field $h$ along [111]  leads to the 
opening of a topological gap $\Delta \sim h^3$ at the Dirac points with Chern numbers ${\cal C} = \pm 1$ of the dispersive Majorana bands. 
 Numerical investigations using exact diagonalisation (ED) and density matrix renormalisation group (DMRG) reported a U(1) gapless intermediate 
phase \cite{Jiang+2019,Pradhan+2020,Hickey+2019} sandwiched between the gapped non-Abelian QSL at small fields and the topologically trivial fully polarised 
phase at large fields. On the other hand, mean-field investigations \cite{Yilmaz+2022,Ralko+2020} and a variational approach \cite{Zhang+2022} found this intermediate 
phase to be gapped. It exhibits ring-like low-energy excitations that might be mistaken for a spinor 
Fermi-surface in numerical studies due to finite-size effects \cite{Zhang+2022}. While DMRG could in principle detect gapless modes through the 
scaling behaviour of the entanglement entropy, no convergence was found in the intermediate phase up to the largest system sizes currently 
accessible \cite{Gohlke+18}.

Here we focus on the first topological  transition, between the two gapped QSL phases. At this transition the total Chern number of the positive bands 
was found to change from ${\cal C}_\textrm{tot}=-1$ in the QSL at small field to ${\cal C}_\textrm{tot}=2$ in the intermediate phase  \cite{Yilmaz+2022,Ralko+2020,Zhang+2022}. 
Such a change in Chern number by $\Delta {\cal C} = 3$ is consistent with a gap closing at the three $M$ points in the Brillouin zone, as indeed observed in previous 
mean-field studies \cite{Yilmaz+2022,Ralko+2020}.
 
The presence of an odd number of Dirac cones at the topological phase transition is unusual and suggests the absence of fermion-doubling as described by the Nielson-Ninomiya theorem \cite{NIELSEN+1981_1,NIELSEN+1981_2,NIELSEN+1981_3,suzuki+2004}. In order to circumvent fermion-doubling it is generically necessary to break at least one of the properties of the Hamiltonian among locality, hermiticity, periodicity, bi-linearity and chiral symmetry. One way is to construct  SLAC fermions \cite{Drell+1976} by introducing long-range hopping terms, which results in a single Dirac cone with singularities at the Brillouin zone boundary \cite{Lang+2019,Liao+2023,Wellegehausen+2017,Wang+2023}. Alternatively, one can break the chiral symmetry explicitly \cite{Wilson+1974,Ginsparg+1982,KAPLAN+1992,NARAYANAN+1995}. For example, in the Bernevig-Hughes-Zhang model \cite{Bernevig+2006} and Qi-Wu-Zhang model \cite{Qi+2006}, a Wilson term \cite{Wilson+1974} acting like a momentum-dependent mass is applied to obtain a single Dirac cone at zero energy. The other Dirac cone is gapped out and pushed into the UV regime. As we will show, the gap closing at the three $M$ points at the field-driven topological phase transition of the Kitaev model is possible because of a UV compensation of Berry curvature due to ring-like hybridisation gaps between the low-energy band and higher-energy Majorana fermion modes.

In order to understand the nature of the topological phase transition we first briefly revisit the Majorana-fermion parton mean-field theory. Our results 
 are in quantitative agreement with those of earlier mean-field studies \cite{Yilmaz+2022,Ralko+2020}. In addition,  we carefully 
 analyse the role of hybridisation gaps and the redistribution of Berry curvature from the UV to the infrared, thereby providing an understanding of why a gap closing 
 at three $M$ points is possible. We further compute the edge states from an armchair ribbon on the two sides of the topological phase transition.
For $h < h_c$, one chiral mode exists on each edge, with chiral central charge of $-1/2$, indicative of non-Abelian anyons and 
consistent with the total Chern number ${\cal C}_\textrm{tot}=-1$ for the positive energy bands
For $h>h_c$, ${\cal C}_\textrm{tot}=2$ corresponds to two chiral Majorana modes
per edge, with central charge $+1$. According to Kitaev’s 16-fold way \cite{Kitaev+2006} the topological order can be described
by a chiral U(1) theory with Abelian semion excitations. Therefore the ground state changes from a gapped $Z_2$
spin liquid to an Abelian $U(1)$ chiral spin liquid across $h_c$.
 
We then use our mean-field results to identify the effective field theory at the topological phase transition around the three $M$ points and 
perform a RG calculation within the Gross-Neveu-Yukawa (GNY) approach that is controlled by the number of fermion flavours $N$. We account for fluctuations of the QSL bond-operators for both intra-valley and inter-valley channels and derive the quantum critical exponents to order $1/N$. We show that inter-valley fluctuations explicitly break Lorentz invariance, resulting in a dynamical exponent that departs from $z=1$. This  suggests that the intermediate topological phase transition of the Kitaev QSL at high field is not in the standard Ising GNY universality class. 
 
 The paper is organised in the following way: in Sec.~\ref{sec.model} we describe the AFM Kitaev model on the honeycomb lattice in 
a magnetic field along [111]. In Sec.~\ref{sec.mf}  we perform a non-perturbative mean-field calculation at finite magnetic field and explicitly 
 calculate the Chern numbers of individual bands, the distribution of Berry curvature in the Brillouin zone, and the edge-state spectra in a strip geometry.
 In Sec.~\ref{sec.FT} we derive the effective field theory of the topological phase transition and carry out the RG analysis of the transition in the presence of 
 fluctuation fields. We derive the critical exponents and characterise the universality class of the transition. In Sec.~\ref{sec.conclude} we summarise our results.

%%%%%%%%%%%%%%

\section{Model}
\label{sec.model}

\begin{figure}[t]
 \includegraphics[width=\linewidth]{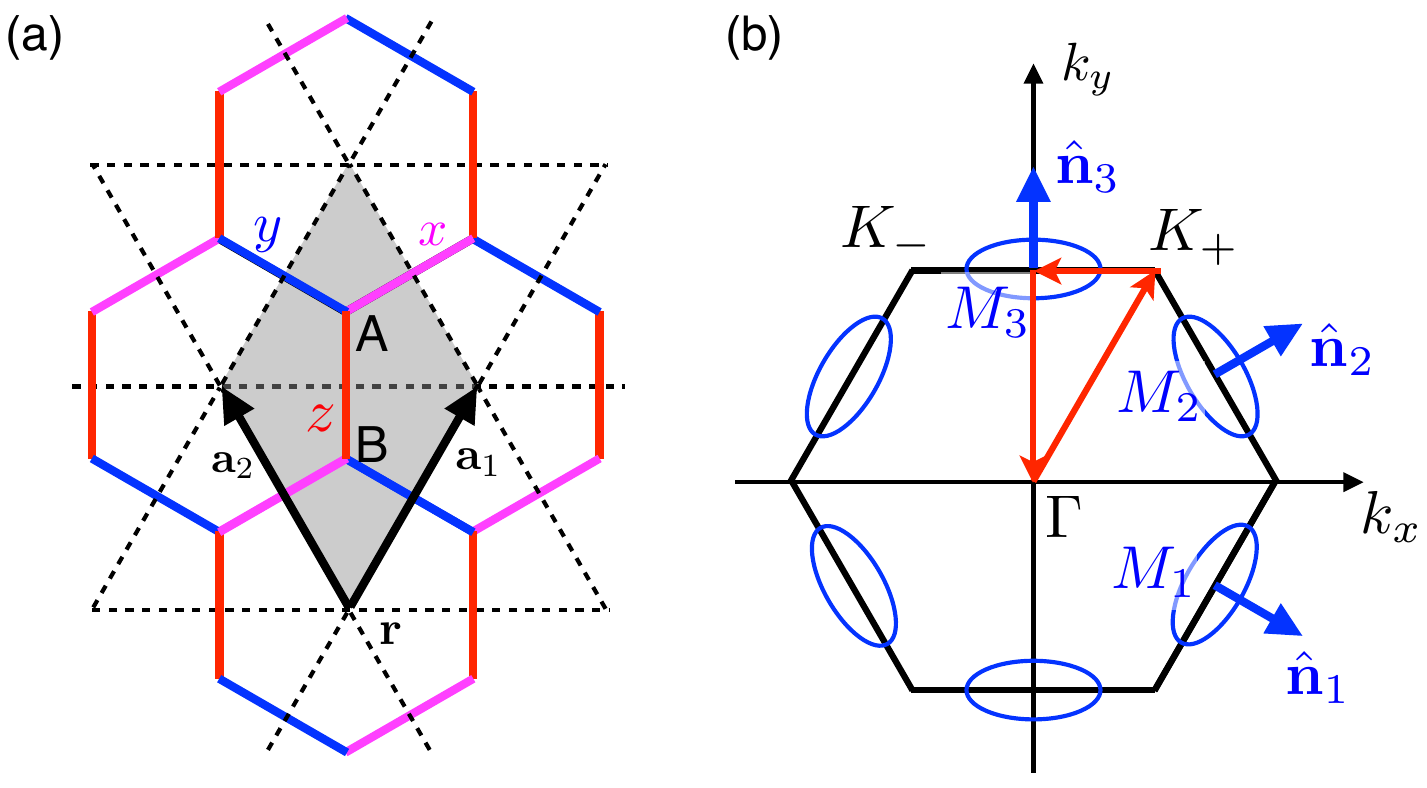}
 \caption{(a) Illustration of the Kitaev model on the honeycomb lattice. The different bonds are labelled by $\gamma=x,y,z$ and shown in different colours. Along a bond of type $\gamma$ only the 
 $\gamma$ components of the spins are coupled. The unit cell is spanned by the lattice vectors $\ba_1$ and $\ba_2$ and contains two sites of the honeycomb lattice, labelled $A$ 
 and $B$. (b) High symmetry points in the hexagonal Brillouin zone. For small $[111]$ field $h$ a topological gap $\Delta\sim h^3$ opens at the two $K$ points. At larger field $h=h_c$ we observe a topological phase transition
 with gap closing at the three $M$ points. The unit vectors $\hat \bn_i$ parametrise the local coordinate frames at $M_i$.}
\label{figure1}
\end{figure}

We consider the AFM  Kitaev model on the honeycomb lattice, which is illustrated in Fig.~\ref{figure1}(a). The key feature of this model is the bond-directional Ising exchange where 
along each of the three different bonds of the honeycomb lattice, labelled by $\gamma=x,y,z$,  only the $\gamma$ components of the spin-1/2 operators are coupled. In addition, the spins are subject
to a magnetic field along the $[111]$ direction. The corresponding Hamiltonian can be written as
\begin{equation}
\label{eq.Ham_spin}
\hat {\cal H} = \sum_{\br,\gamma=x,y,z} \Big\{ K \hat \sigma_a^\gamma(\br) \hat \sigma_b^\gamma(\br+\bd_\gamma) - \frac{h}{\sqrt{3}} \sum_{s=a,b}\hat\sigma_s^\gamma(\br)\Big\},
\end{equation}
where $\hat\sigma^\gamma_s(\br)$ are spin-1/2 operators on the sites $s=a,b$ in the unit cell $\br$ of the triangular lattice spanned by $\ba_{1,2}=(\pm \sqrt{3}/2, 3/2)$. Using these conventions, the 
lattice vectors for the three different bonds are given by $\bd_x=\ba_1$, $\bd_y=\ba_2$, and $\bd_z=\bm{0}$.  Measuring the spins in units  
of $\hbar/2$, the spin commutator relations read  $[\hat\sigma_s^\alpha(\br),\hat\sigma_{s'}^\beta(\br')] = 2 \delta_{\br,\br'}\delta_{s,s'}\epsilon_{\alpha\beta\gamma}\hat\sigma_s^\gamma(\br)$. 

In the absence of field, $h=0$, the Kitaev model is exactly solvable by expressing the spin operators in terms of a set of four Majorana fermion operators $\hat\eta_s^\mu(\br)$ ($\mu=0,x,y,z$),
\begin{equation}
\hat \sigma_s^\gamma(\br) = i \hat\eta^0_s(\br) \hat \eta^\gamma_s(\br).
\end{equation}
 The Majorana fermions satisfy the Clifford algebra $\{\hat \eta_s^\mu(\br), \hat \eta_{s'}^\nu(\br') \} = 2 \delta_{\br,\br'} \delta_{s,s'} \delta_{\mu,\nu}$ and are subject to the local constraints
 \begin{equation}
 \hat\eta^0_s(\br) \hat \eta^\gamma_s(\br) + \frac{1}{2}\epsilon_{\alpha\beta\gamma} \hat \eta_s^\alpha(\br) \hat \eta_s^\beta(\br) = 0,
 \label{eq.constraint}
 \end{equation}
 in order to correctly represent the spin Hilbert space. 
 
\begin{figure*}[t]
 \includegraphics[width=\linewidth]{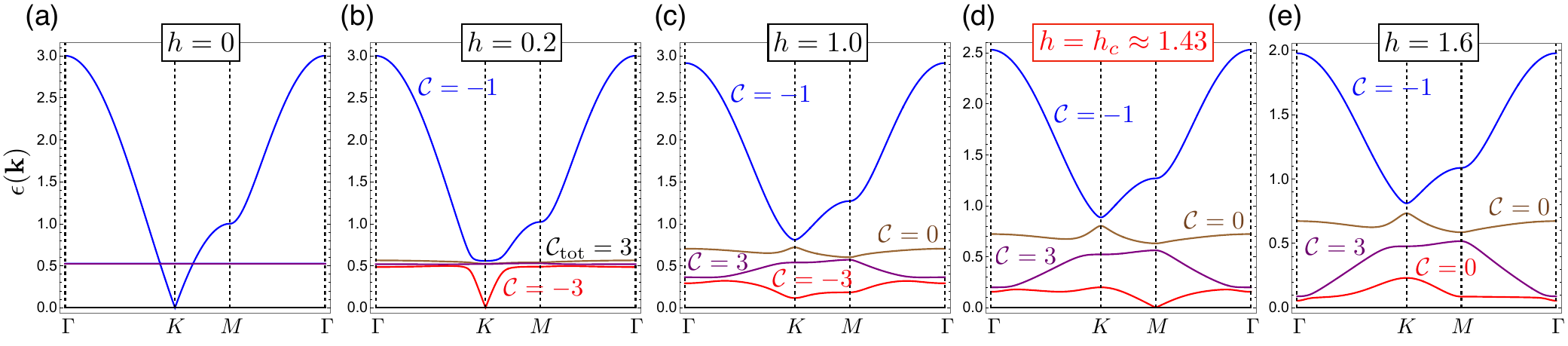}
 \caption{Evolution of the Majorana fermion mean-field spectrum along a high-symmetry path as a function of magnetic field $h$ along $[111]$. (a) Spectrum of the Kitaev model at $h=0$, showing a dispersive Majorana band with Dirac 
 cones at the $K$ points and three degenerate flat bands. (b) For small $h$ a small gap $\Delta\sim h^3$ open at the $K$ points and the dispersive band hybridizes with the flat ones. All bands acquire non-zero Chern numbers, where ${\cal C}=-3$ for the 
 low-energy band. (c) At $h=1$ a softening at the $M$ points becomes visible. The Chern numbers remain unchanged.  (d) At the critical point $h = h_c\approx 1.43$ the gap closes at the three $M$ points. 
 (e) Bands at $h=1.6$. The Chern number of the low-energy band changes to ${\cal C}=0$ at $h>h_c\approx 1.43$, indicating  
the presence of a topological phase transition. }
\label{figure2}
\end{figure*}

 Although the Kitaev model is quartic in terms of the Majorana fermions, an exact solution can be obtained because the bond operators 
 $\hat A_\gamma(\br) = i \hat\eta_a^\gamma(\br) \hat\eta_b^\gamma(\br+\bd_\gamma)$, which have eigenvalues $\pm 1$,  are local and commute with the Hamiltonian. As a consequence, it is sufficient to diagonalise 
 the quadratic Hamiltonian for the $\hat \eta^0$ Majorana fermion for a given realisation of fluxes, which are obtained by multiplying the corresponding bond operator eigenvalues around each plaquette. In the zero-flux 
 ground-state sector this results in the energy dispersion $\pm K |1+e^{i\bk \cdot \ba_1}+e^{i\bk \cdot \ba_2}|$ for $\hat \eta^0$, with Dirac points at the corners $K_\pm$ of the 
 hexagonal Brillouin zone [see Fig.~\ref{figure1}(b)],  
 and three degenerate flat bands of the local $\hat \eta^x$, $\hat \eta^y$, and  $\hat \eta^z$ Majorana fermions, shown in Fig.~\ref{figure2}(a).

%%%%%%%%%%%%%%

\section{Mean-field theory}
\label{sec.mf}

A magnetic field along the $[111]$ direction represents the simplest way to break the exact solvability of the Kitaev model. For $h>0$ the different Majorana fermions 
hybridise and the plaquette operators no longer commute with the Hamiltonian. As a first step we use self-consistent mean-field theory to decouple the quartic Majorana fermion Hamiltonian, introducing the bond expectation values  
\begin{eqnarray}
A & = & \langle i \hat\eta_a^\gamma(\br) \hat\eta_b^\gamma(\br+\bd_\gamma)\rangle,\nn\\
B & = & \langle i \hat\eta_a^0(\br) \hat\eta_b^0(\br+\bd_\gamma)\rangle,
\end{eqnarray}
which by symmetry take the same values on all nearest-neighbour bonds. To account for effects of internal magnetic fields we also 
decouple in the local magnetisation channel,  
\begin{equation}
m_\gamma = \frac{m}{\sqrt{3}} = \langle i  \hat\eta^0_s(\br) \hat \eta^\gamma_s(\br) \rangle.
\end{equation}
We treat the three local constraints (\ref{eq.constraint}) on average through Lagrange multipliers $\lambda_\gamma = \lambda/\sqrt{3}$, 
with the additional 
contribution to the Hamiltonian, 
\begin{equation}
\hat {\cal H}_\lambda = i K \frac{\lambda}{\sqrt{3}} \sum_{\br,s,\gamma}  \Big\{ \hat\eta^0_s(\br) \hat \eta^\gamma_s(\br) + \frac{1}{2}\epsilon_{\alpha\beta\gamma} \hat \eta_s^\alpha(\br) \hat \eta_s^\beta(\br)\Big\}.
\end{equation}

In momentum space the mean-field Hamiltonian has an 8-by-8 matrix structure ($\mu=0,x,y,z$ and $s=a,b$). In the following, we measure all energies 
in units of the Kitaev coupling $K$. For given mean-field parameters $A, B, m$ and Lagrange parameter $\lambda$ we can numerically diagonalise this matrix 
for each momentum $\bk$ in the Brillouin zone, resulting in 4 pairs of energy eigenvalues 
$\pm \epsilon_i(\bk,A,B,m,\lambda)$ with $\epsilon_i\ge 0$ and a zero-temperature mean-field energy per unit cell
\begin{equation}
E = -\sum_{i=1}^4 \int_{\rm BZ} \frac{d^2\bk}{V_{\rm BZ}} \epsilon_i(\bk,A,B,m,\lambda) +3 AB +m^2,
\end{equation} 
where $V_{\rm BZ}$ stands for the area of the Brillouin zone. 
The three equations $\partial E/\partial x= 0$ with $x=A,B,m$ can be conveniently solved using the standard iterative procedure. However, at each iteration 
step we need to determine the Lagrange multiplier $\lambda$ 
by solving the integral equation $\partial E/\partial \lambda = 0$ with bisection.

\subsection{Mean-field results}

For $h=0$ the mean-field equations can be solved analytically and reproduce the exact solution of the Kitaev 
model, shown in Fig. 2a. 
For small $h$ a very small gap opens at the $K$ points, which is still barely visible at $h=0.2$ (see Fig.~\ref{figure2}(b)). In addition, the hybridisation between t
he $\hat\eta^0$ and $\hat\eta^\gamma$ Majorana fermions results in a ring like gap feature around the $K$ points. All bands are topologically non-trivial and carry 
non-zero Chern numbers. Previously, it was demonstrated that the $[111]$ field leads to a topological gap opening of the $\hat\eta^0$ Dirac mode in third order perturbation 
theory \cite{Kitaev+2006}. Ignoring hybridisation effects this then results in Chern numbers ${\cal C}=\pm 1$ of the gapped low energy $\hat\eta^0$ mode. 
While the Chern numbers of all four positive bands still add up to ${\cal C}_\textrm{tot}=-1$, the formation of hybridisation gaps is responsible for a redistribution of Chern numbers between bands, resulting in ${\cal C}=-3$ for the 
low energy band.

\begin{figure*}
	\includegraphics[width=0.6\linewidth]{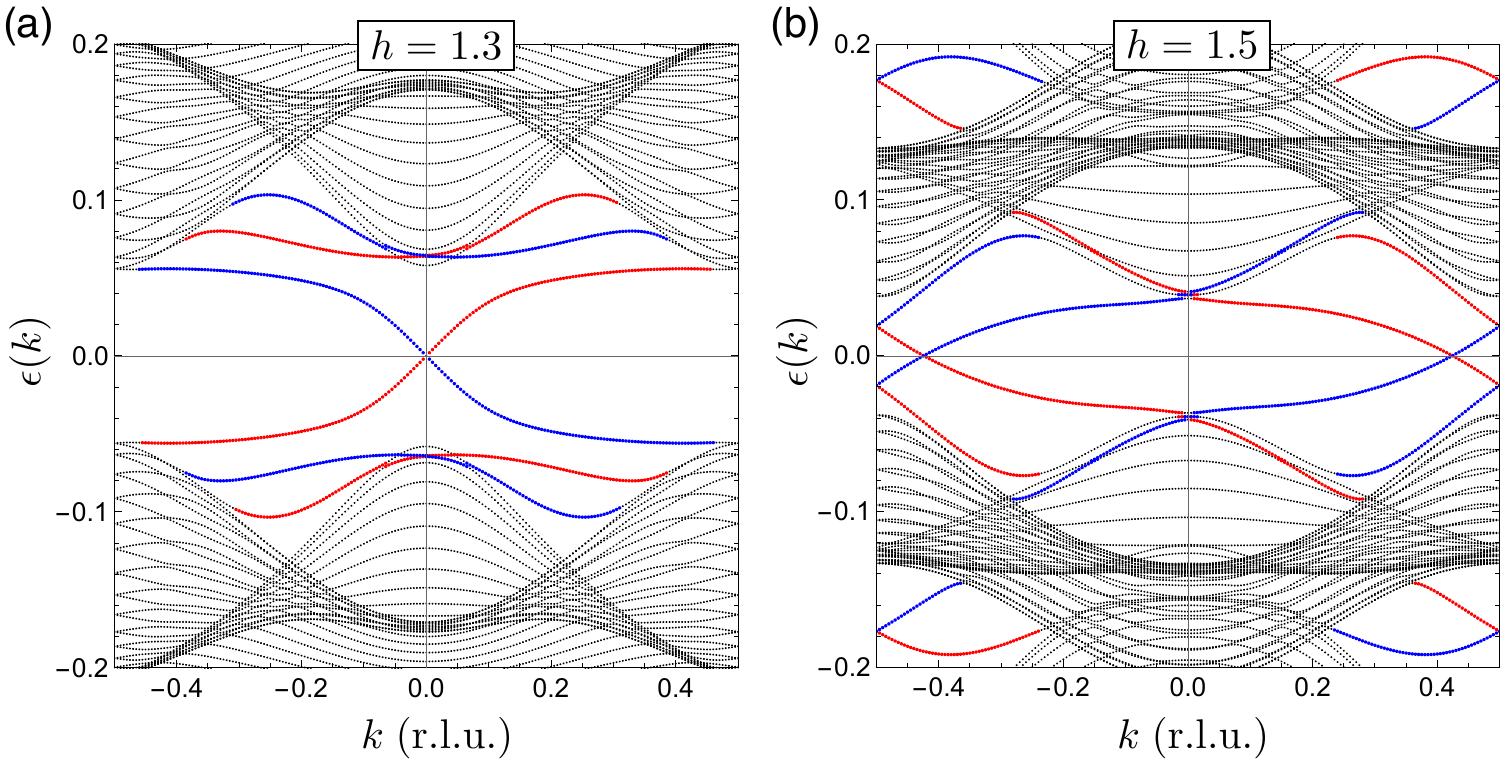}
	\caption{Edge states calculated from an armchair ribbon at (a) $h=1.3$ and (b) $h=1.5$. Bands in blue (red) correspond to the edge states at the left (right) edge. There is one chiral mode per edge for $h<h_c$, and two chiral modes per edge for $h>h_c$, corresponding to the change of total Chern number from $-1$ to $+2$.}
	\label{figure3}
\end{figure*}

Increasing the field further to $h=1$ the gap at the $K$ points increases and a softening of the dispersion at the $M$ points is observed (see Fig.~\ref{figure2}(c)). At the critical field $h_c\approx 1.43$, the gap closes at the three $M$ points as shown in Fig.~\ref{figure2}(d), which have massless Dirac low energy quasiparticles. As pointed out earlier, an odd number of Dirac points  is  unusual and only possible if additional Berry curvature is located elsewhere in the Brillouin zone. We address in detail the evolution of the distribution of Berry curvature in the low energy band in Appendix A. 

Beyond  the critical field, $h>h_c\approx 1.43$,  the gap reopens and the Chern number of the low energy band  changes to ${\cal C}=0$ (see Fig.~\ref{figure2}(e)), indicating the presence of a topological phase transition of the low energy band at $h=h_c$. This results in a total Chern number for the positive energy bands $\mathcal{C}_\mathrm{tot} = +2$. This critical field value and the total Chern numbers around the phase transition are consistent with values reported in the literature \cite{Yilmaz+2022,Ralko+2020}, noting that since we measure spins in units of $\hbar/2$ our field values are rescaled  by a factor of two. 

The change in total Chern number reflects a change in the number of chiral Majorana edge states due to the bulk-boundary correspondence. The edge modes of an armchair ribbon at $h=1.3$ and $h=1.5$ are shown in Fig.~\ref{figure3}. The converged mean-field parameters of the Hamiltonian are used in the calculation. For $h<h_c$, one chiral mode exists on each edge, with chiral central charge of $-1/2$, indicative of non-Abelian anyons with topological spin $-\pi/8$. For $h>h_c$, the total Chern number ${\cal C}_{\mathrm{tot}}=+2$ corresponds to two  chiral Majorana modes per edge, with central charge $+1$. According to Kitaev's 16-fold way \cite{Kitaev+2006}, the topological order can be described by a chiral $U(1)$ theory with Abelian semion excitations. Therefore the ground state changes from a gapped $Z_2$ spin liquid to an Abelian $U(1)$ chiral spin liquid across $h_c$.

\section{Field Theoretical Analysis of Topological Phase Transition}
\label{sec.FT}

\subsection{Effective Field Theory}

At the topological phase transition the gap closes at the three $M$ points in the Brillouin zone, which we will label by $M_i$ ($i=1,2,3$) in the following, 
as illustrated in Fig.~\ref{figure1}(b). In a small momentum region 
around these points and for fields $h$ close to $h_c$ the low energy Hamiltonian around $M_i$ will have the  conventional 2-by-2 spinor matrix structure of a gapped Dirac point, where the gap $\Delta\sim (h-h_c)$ is the  same at all $M$ points. By symmetry, one would further expect that the Hamiltonians $\hat{\cal H}_i$ at each $M_i$ are identical when expressed in the local coordinate frame relative to the edge of the Brillouin zone with normal 
vector $\hat{\bn}_i$ (Fig.~\ref{figure1}). The Hamiltonian matrices should therefore take the form
\begin{equation}
\mathbf{H}_i (\bk)=  v_\parallel [(\hat{\bn}_i\times \ez)\cdot\bk ]\bm{\tau}_x + v_\perp(\hat{\bn}_i\cdot\bk)\bm{\tau}_y+\Delta \bm{\tau}_z,
\label{Ham1}
\end{equation}
where $\bm{\tau}_\gamma$ are Pauli matrices in pseudo-spin space. We confirmed this expected form  numerically by projecting the full mean-field Hamiltonian onto the low-energy sector at each $M_i$ and treating $(h-h_c)$ and the momentum shifts $k_x$, $k_y$ away from $M_i$ as small perturbations. 

Our numerical results show a small anisotropy of Fermi velocities, $(v_\parallel-v_\perp)/v_\perp\approx 0.05$. In the following we will neglect this anisotropy and absorb the velocity $v=v_\parallel=v_\perp$ in a re-definition of 
$\bk$. Our RG analysis will indeed confirm that $v_\parallel=v_\perp$ at the critical fixed point. 

The three Dirac Hamiltonians shown in Eq. (\ref{Ham1}) have the topological charge ${\rm sgn}(\Delta)/2$.  The 
additional topological charge of $-3/2$ associated with the nodal line gap around $\Gamma$ is located at top of the low-energy band and  remains unchanged across the transition. This feature  has no direct effect on the 
nature of the topological phase transition, other than permitting the emergence of an odd number of Dirac cones in the IR via UV compensation. In the following we focus on the bulk properties in the thermodynamic limit, in the absence of zero energy edge modes.  

Writing the partition function as a Grassmann path-integral over Majorana fermion fields, the low-energy free fermion action at the critical point ($\Delta=0$) is given by
\begin{eqnarray}
\mathcal{S}_0[\bar{\bm{\psi}},\bm{\psi}] & = &  \sum_{i=1}^3\sum_{\nu=1}^N \int_{\vec{k}=(k_0,\bk)}\bar{\bm{\psi}}_{i\nu}(\vec{k})\Big\{-ik_0 \nn\\
& & +[(\hat{\bn}_i\times \ez)\cdot\bk ]\bm{\tau}_x + (\hat{\bn}_i\cdot\bk)\bm{\tau}_y\Big\}\bm{\psi}_{i\nu}(\vec{k}), 
\end{eqnarray}
where we have introduced the frequency momentum 3-vector $\vec{k}=(k_0,\bk)$, $i=1,2,3$ labels the three $M$-point Dirac valleys, and we have generalised to $\nu=1,\ldots,N$ replicas of the theory, enabling a systematic expansion 
in $1/(3N)$. The resulting fermion Green function at $M_i$ is given by
\begin{equation}
\bG_i(\vec{k}) = \frac{ik_0 +[(\hat{\bn}_i\times \ez)\cdot\bk ]\bm{\tau}_x + (\hat{\bn}_i\cdot\bk)\bm{\tau}_y}{{\vec{k}}^2}.
\label{eq.fermionGreen}
\end{equation}

To understand the critical behaviour we need to include fluctuations beyond mean-field theory. The starting point would be a Hubbard-Stratonovich decoupling of the interaction terms in the initial lattice model, which, at saddle-point level, 
reproduces the mean-field theory. While the initial fluctuation fields couple to bond operators $i \hat\eta_a^\gamma(\br) \hat\eta_b^\gamma(\br+\bd_\gamma)$ and $i \hat\eta_a^0(\br) \hat\eta_b^0(\br+\bd_\gamma)$, which lead to a Yukawa coupling in the sublattice channel $\bm{\tau}_y$ as discussed in \cite{Hu+2024}, here the Majorana fermions of different flavours and on different sub-lattices mix under the unitary transformation that diagonalises the mean-field Hamiltonian. We will therefore obtain fluctuations in all channels $\bm{\tau}_\gamma$ of the new low-energy pseudo-spin
space. In general, most fluctuations will be gapped and can be neglected. One exception is when the system  is close to a symmetry-breaking instability, in which case he corresponding fluctuations can become soft. Here we won't study such multi-criticality and only 
focus on the dynamical fluctuations of the mass gap $\Delta$ in the $\bm{\tau}_z$ channel, since it is the tuning parameter of the topological phase transition at the quantum critical point ($\Delta=0$).  

Since the interactions are short ranged, both inter- and intra-valley fluctuations will be important. On the full two-dimensional Brillouin zone the fluctuations in the mass channel are of the form 
$\Phi(\bQ)\bar{\bm{\psi}}(\bK)\bm{\tau}_z \bm{\psi}(\bK+\bQ)$, where we have dropped the dependence on frequencies, for brevity.  In the low-energy theory we only consider momentum patches close to the $M$ points, 
$\bK=\bM_i+\bk$, and define  $\bm{\psi}(\bK) = \bm{\psi}(\bM_i+\bk)\equiv \bm{\psi}_i(\bk)$. 
For the  momentum transfer we write $\bQ=\bQ_{ij}+\bq$ where $\bQ_{ij} = \bM_j-\bM_i$ and $\bq$ is small. Defining the intra-valley fluctuation fields ($i=j$) as $\Phi(\bQ_{ii}+\bq)=\Phi(\bq)=\phi(\bq)$ and the 
inter-valley fluctuation fields ($i\neq j$) as $\Phi(\bQ_{ij}+\bq)=\varphi_{ij}(\bq)$, the Yukawa couplings can be written as
\begin{eqnarray}
\mathcal{S}_Y & = &  \frac{g}{\sqrt{3N}} \sum_i \sum_\nu \int_{\vec{k},\vec{q}}\phi(\vec{q}) \bar{\bm{\psi}}_{i\nu}(\vec{k})\bm{\tau}_z \bm{\psi}_{i\nu}(\vec{k}+\vec{q})\\
& & +  \frac{\tilde{g}}{\sqrt{6N}} \sum_{i,j}^{i\neq j} \sum_\nu \int_{\vec{k},\vec{q}}\varphi_{ij}(\vec{q}) \bar{\bm{\psi}}_{i\nu}(\vec{k})\bm{\tau}_z \bm{\psi}_{j\nu}(\vec{k}+\vec{q}),\nn
\end{eqnarray}
where $g$ and $\tilde{g}$ are the strengths of the intra- and inter-valley Yukawa couplings, respectively. Note that $\varphi^*_{ij}(\vec{q}) = \varphi_{ji}(-\vec{q})$. Finally, the quadratic actions for the fluctuation fields are given by
\begin{eqnarray}
\mathcal{S}_0[\phi] & = & \frac12 \int_{\vec{q}} D^{-1}(\vec{q}) |\phi(\vec{q})|^2\\
\mathcal{S}_0[\varphi] & = & \frac{1}{2} \sum_{i,j}^{i< j} \int_{\vec{q}} \tilde{D}^{-1}(\vec{q}) |\varphi_{ij}(\vec{q})|^2,
\end{eqnarray}
where the inverse boson propagators are given by the usual quadratic gradient and mass terms and a self energy correction, $D^{-1}(\vec{q}) = q_0^2+c^2\bq^2 +m^2+\Pi(\vec{q})$ and 
$\tilde{D}^{-1}(\vec{q}) = q_0^2+\tilde{c}^2\bq^2 +\tilde{m}^2+\tilde{\Pi}(\vec{q})$.

\subsection{Self-energy corrections and IR boson propagators}

\begin{figure}[t]
 \includegraphics[width=\linewidth]{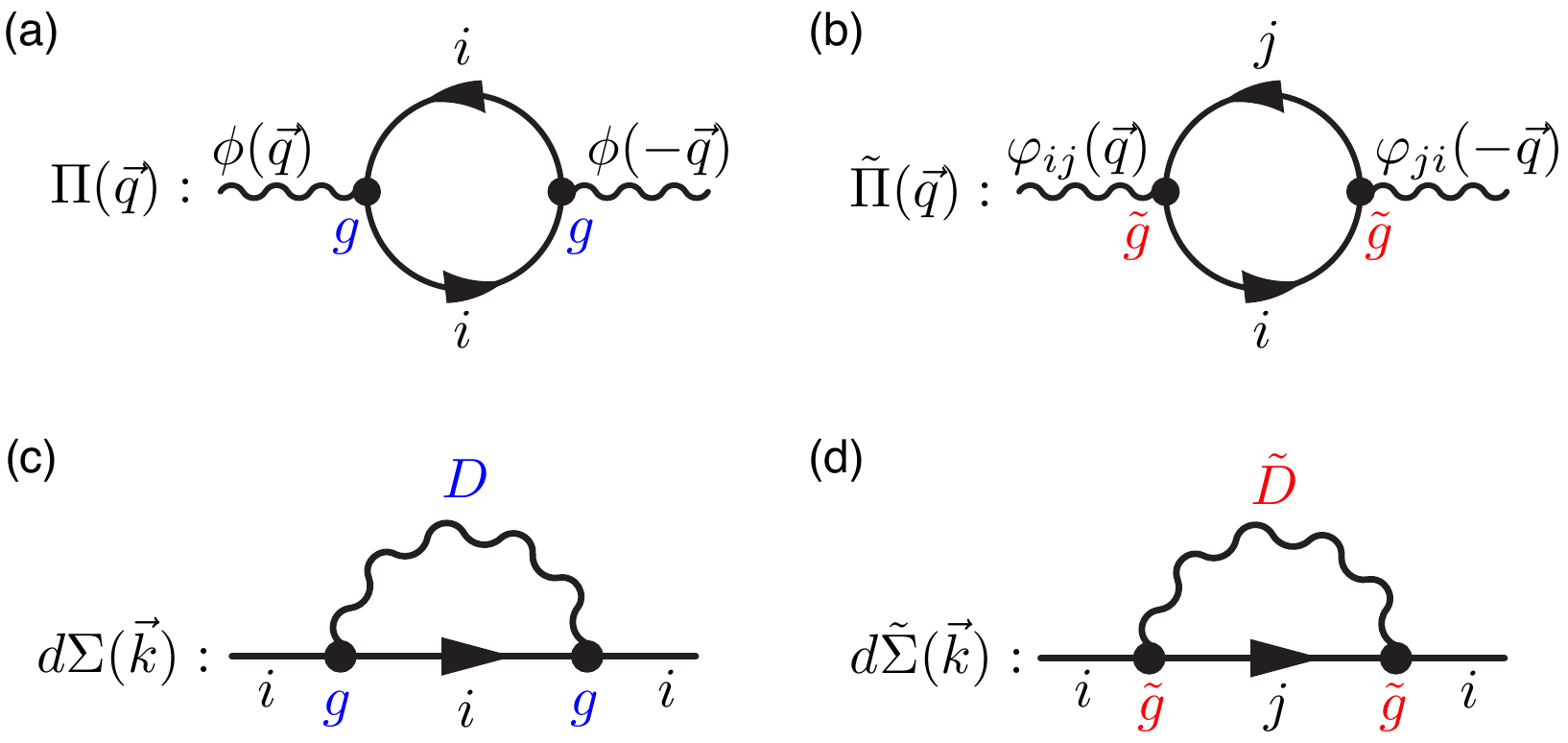}
 \caption{Self-energy diagrams of the GNY model. Top row: Bosonic self-energy diagrams. Fermionic polarisation bubble diagrams give non-analytic IR corrections to boson propagators of (a) intra-valley and (b) inter-valley fluctuation fields ($i\neq j$). Bottom row: fermionic self energy corrections, leading to a renormalisation of the free fermion action. (c) shows the contribution $d\Sigma(\vec{k})$ from intra-valley fluctuations, (d) the contribution 
 $d\tilde{\Sigma}(\vec{k})$ from fluctuations between different valleys, $i\neq j$.}
\label{figure4}
\end{figure}

The self-energy corrections to the boson propagators correspond to the diagrams shown in Figs.~\ref{figure4} (a) and (b) and are given by $\Pi(\vec{q}) = g^2 [f_{ii}(\vec{q})- f_{ii}(\vec{0})]$ and $\tilde{\Pi}(\vec{q}) = \frac{\tilde{g}^2}{3} [f_{ij}(\vec{q})- f_{ij}(\vec{0})]$ 
for $i\neq j$, where
\begin{equation}
f_{ij} (\vec{q}) = \int_{\vec{k}} \textrm{Tr} \Big\{\bG_i(\vec{k})\bm{\tau}_z \bG_j(\vec{k}+\vec{q})\bm{\tau}_z   \Big\}.
\label{eq.selfenergy0}
\end{equation}

The polarisation diagrams can be computed analytically, see Appendix \ref{app.a}, resulting in 
\begin{equation}
f_{ij} (\vec{q})-f_{ij} (\vec{0}) = \frac{2(1+\cos\beta_{ij})q_0^2+(1+3\cos\beta_{ij})\bq^2}{32|\vec{q}|},
\label{eq.selfenergy}
\end{equation}
where $\beta_{ij}$ is the angle between the unit vectors $\hat{\bn}_i$ and $\hat{\bn}_j$ at valleys $M_i$ and $M_j$, $\cos\beta_{ij}=\hat{\bn}_i\cdot\hat{\bn}_j$.
Using that $\cos\beta_{ij}=1$ for the intra-valley ($i=j$) and  $\cos\beta_{ij}=\cos(\pm\pi/3)=1/2$ for the inter-valley terms ($i\neq j$), we obtain the boson self-energy corrections
\begin{eqnarray}
\Pi(\vec{q}) & = & \frac{g^2}{8} |\vec{q}|, \label{intra_Landau_daming} \\
\tilde{\Pi}(\vec{q}) & = & \frac{\tilde{g}^2}{32} |\vec{q}|\left(1-\frac16 \frac{\bq^2}{\vec{q}^2}   \right),
\end{eqnarray} 
for intra and inter-valley fluctuation channels, respectively. The intra-valley correction (\ref{intra_Landau_daming}) is the standard result as shown in \cite{Uryszek+20}. 

The self-energy corrections of both channels scale as $\sim|\vec{q}|=\sqrt{q_0^2 +\bq^2}$ and hence dominate over the conventional quadratic gradient terms 
in the IR limit. In order to understand the universality of the transition it is sufficient to keep the leading frequency and momentum dependence and use 
$D^{-1}(\vec{q})=\Pi(\vec{q})$ and  $\tilde{D}^{-1}(\vec{q})=\tilde{\Pi}(\vec{q})$ as inverse IR propagators.

\subsection{Renormalisation-group analysis}

\begin{figure}[t]
 \includegraphics[width=0.95\linewidth]{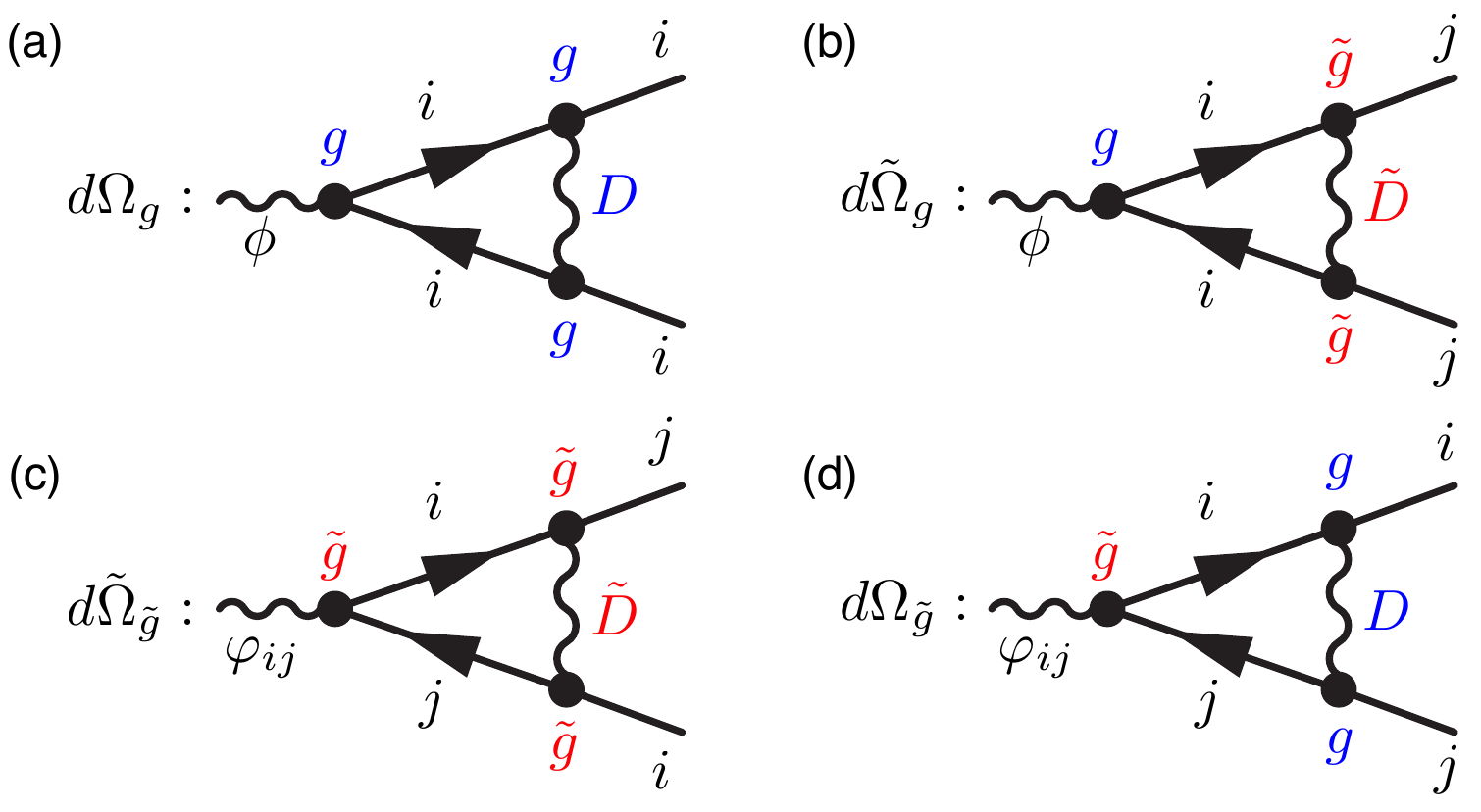}
 \caption{Panels (a) and (b) show the diagrams that renormalise the intra-valley Yukawa coupling $g$, (c) and (d) the diagrams that renormalise the Yukawa coupling $\tilde{g}$ between different valleys $i\neq j$.}
\label{figure5}
\end{figure}

We proceed to analyse the GNY action $\mathcal{S}_0[\bar{\bm{\psi}},\bm{\psi}] + \mathcal{S}_0[\phi] + \mathcal{S}_0[\varphi] +\mathcal{S}_Y[\bar{\bm{\psi}},\bm{\psi},\phi,\varphi]$ using a 
renormalisation group approach. We integrate out UV modes from the infinitesimal three dimensional frequency-momentum shell,
\begin{equation}
\Lambda e^{-d\ell}\le |\vec{k}|\le\Lambda,
\label{eq.shell}
\end{equation}
where $\Lambda$ denotes the original cut-off of the theory. The diagrams that renormalise the free fermion action and the Yukawa couplings are shown in the bottom row of  Fig. \ref{figure4} and  in Fig. \ref{figure5}, respectively. Note that the IR boson 
propagators are non-analytic and do not renormalise under the perturbative shell RG scheme. The shell integration is followed by a rescaling of frequency and momentum, 
\begin{equation}
k_0 \to k_0 e^{-z d\ell}, \;\; \bk\to\bk e^{-d\ell},
\end{equation}
and fields
\begin{equation}
\bm{\psi} \to \bm{\psi} e^{-\Delta_{\bm{\psi}}/2\, d\ell}, \;\; \phi \to \phi e^{-\Delta_\phi/2 \,d\ell}, \;\; \varphi \to \varphi e^{-\Delta_\varphi/2 \,d\ell}.
\end{equation}

We start by analysing the shell corrections to the free fermion action, corresponding to the diagrams in Figs.~\ref{figure4} (c) and (d), and given by 
\begin{equation}
d\mathcal{S}_0[\bar{\bm{\psi}},\bm{\psi}] = \sum_{i,\nu} \int_{\vec{k}}^< \bar{\bm{\psi}}_{i\nu}(\vec{k}) \left[ d\bm{\Sigma}_i(\vec{k}) +d\tilde{\bm{\Sigma}}_i(\vec{k})  \right] \bm{\psi}_{i\nu}(\vec{k}),
\end{equation}
 with
 \begin{eqnarray}
 \label{eq.fermion_ren1}
 d\bm{\Sigma}_i(\vec{k}) & = & -\frac{g^2}{3N} \int_{\vec{q}}^> D(\vec{q})\bm{\tau}_z \bG_i(\vec{k}+\vec{q})\bm{\tau}_z, \\
    \label{eq.fermion_ren2}
  d\tilde{\bm{\Sigma}}_i(\vec{k}) & = & -\frac{\tilde{g}^2}{6N} \sum_{j(\neq i)} \int_{\vec{q}}^> \tilde{D}(\vec{q})\bm{\tau}_z \bG_j(\vec{k}+\vec{q})\bm{\tau}_z,
 \end{eqnarray}
for the infinitesimal intra- and inter-valley fermion self-energy corrections. In the above and the following, ``$>$" denotes integration over the infinitesimal shell (\ref{eq.shell}), ``$<$" over modes up to the
reduced cut-off, $|\vec{k}|\le \Lambda e^{-d\ell}$.

Expanding Eqs. (\ref{eq.fermion_ren1}) and (\ref{eq.fermion_ren2}) to linear order in outer frequency $k_0$ and momenta $\bk$, 
\begin{eqnarray}
\label{eq.fermion_ren3}
d\bm{\Sigma}_i(\vec{k}) +d\tilde{\bm{\Sigma}}_i(\vec{k}) & = & -ik_0 \left(\Sigma_0+\tilde{\Sigma}_0   \right)d\ell \nn\\
& & + [(\hat{\bn}_i\times \ez)\cdot\bk ]\bm{\tau}_x \left( \Sigma_x+\tilde{\Sigma}_x  \right) d\ell \nn\\
& & + (\hat{\bn}_i\cdot\bk)\bm{\tau}_y\left( \Sigma_y+\tilde{\Sigma}_y  \right)d\ell,
\end{eqnarray} 
and carrying out the resulting shell integrals over $\vec{q}$ for the coefficients $\Sigma_n d\ell$ and $\tilde{\Sigma}_n d\ell$ ($n=0,x,y$), we obtain
\begin{equation}
\Sigma_0 = \Sigma_x = \Sigma_y = \frac{1}{3N} \frac{4}{3\pi^2}
\end{equation}
for the intra-valley contributions and 
\begin{equation}
\tilde{\Sigma}_0  =  \frac{1}{3N}\frac{96(11\kappa-2)}{\pi^2}, \quad \tilde{\Sigma}_{x,y} =  \frac{1}{3N}\frac{48(1-5\kappa)}{\pi^2}
\end{equation}
for the inter-valley terms, where we have defined $\kappa=(1/\sqrt{5})\arctan\left(1/\sqrt{5}\right)$. Details of the calculation can be found in Appendix \ref{app.b}. Combining the loop corrections with the rescaling contributions and demanding that the 
inverse fermion propagator remains invariant under RG we obtain the conditions
\begin{eqnarray}
 -2-2z-\Delta_\psi+\Sigma_0+\tilde{\Sigma}_0 & = & 0, \\
-3 -z -\Delta_\psi+\Sigma_{x,y}+\tilde{\Sigma}_{x,y} & = & 0,
\end{eqnarray}
from the scale invariance of the frequency and spatial momentum coefficients, respectively. These equations allow us to determine the dynamical exponent 
\begin{eqnarray}
\label{eq.z}
z & = &  1 + \tilde{\Sigma}_0 - \tilde{\Sigma}_{x,y} \nn\\
& = & 1+\frac{1}{3N} \frac{48(27\kappa-5)}{\pi^2} \approx 1+\frac{0.379}{3N},
\end{eqnarray}
and the fermion anomalous dimension
\begin{eqnarray}
\label{eq.etapsi}
\eta_\psi & = &  \Sigma_0 -\tilde{\Sigma}_0+2\tilde{\Sigma}_{x,y}\nn\\
& = & \frac{1}{3N} \frac{4(217-1152\kappa)}{3\pi^2} \approx \frac{0.047}{3N}
\end{eqnarray}
which measures the deviation of $\Delta_\psi$ from tree-level scaling, $\Delta_\psi =-4+\eta_\psi$.

Finally, we compute the renormalisation of the Yukawa couplings $g$ and $\tilde{g}$. The corresponding one-loop  diagrams involve contractions of three Yukawa vertices and are shown in Fig.~\ref{figure5}. Note that 
since $g^2 D(\vec{q}) \sim g^0$ and  $\tilde{g}^2 \tilde{D}(\vec{q}) \sim \tilde{g}^0$ the RG equations for $g$ and $\tilde{g}$ decouple and take the simple linear form 
\begin{eqnarray}
\label{eq.dg}
\frac{dg}{d\ell} & = & \left(-4-2 z-\Delta_\psi-\frac{\Delta_\phi}{2}+\Omega_g +\tilde{\Omega}_g  \right)g, \\
\frac{d\tilde{g}}{d\ell} & = & \left( -4-2 z-\Delta_\psi-\frac{\Delta_\varphi}{2}+\Omega_{\tilde{g}} +\tilde{\Omega}_{\tilde{g}}    \right)\tilde{g}, 
\label{eq.dgt}
\end{eqnarray}
where we have included the rescaling contributions. The one-loop corrections $d\Omega_g = \Omega_g d\ell$ and $d\tilde{\Omega}_g = \tilde{\Omega}_g d\ell$ that contribute 
to the renormalisation of the intra-valley coupling $g$ are given by the shell integrals
\begin{eqnarray}
\label{eq.Yuk1}
\Omega_g d\ell & = & \frac{g^2}{3N} \int_{\vec{q}}^> D(\vec{q}) \bG_i(\vec{q}) \bm{\tau}_z \bG_i(\vec{q}) \bm{\tau}_z  =  - \frac{d\ell}{3N}\frac{4}{\pi^2}, \\
\label{eq.Yuk2}
\tilde{\Omega}_g d\ell & = & \frac{\tilde{g}^2}{3N} \int_{\vec{q}}^> \tilde{D}(\vec{q}) \bG_i(\vec{q}) \bm{\tau}_z \bG_i(\vec{q}) \bm{\tau}_z  =  -\frac{d\ell}{3N}\frac{96\kappa}{\pi^2},\quad
\end{eqnarray}  
which correspond to the diagrams in Fig.~\ref{figure5}(a) and (b). In the same way, the diagrammatic contributions that contribute to the renormalisation of the inter-valley Yukawa coupling $\tilde{g}$ are given by
\begin{eqnarray}
\label{eq.Yuk3}
\Omega_{\tilde{g}} d\ell & = & \frac{g^2}{3N} \int_{\vec{q}}^> D(\vec{q}) \bG_i(\vec{q}) \bm{\tau}_z \bG_j(\vec{q}) \bm{\tau}_z  = - \frac{d\ell}{3N}\frac{8}{3 \pi^2},\quad\\
\label{eq.Yuk4}
\tilde{\Omega}_{\tilde{g}} d\ell & = & \frac{\tilde{g}^2}{6N} \int_{\vec{q}}^> \tilde{D}(\vec{q}) \bG_i(\vec{q}) \bm{\tau}_z \bG_j(\vec{q}) \bm{\tau}_z \nn\\
& = &   -\frac{d\ell}{3N}\frac{24(1-4\kappa)}{\pi^2},
\end{eqnarray}  
and correspond to the diagrams in Fig.~\ref{figure5}(c) and (d). Detail on the calculation of the above one-loop diagrams can be found in Appendix \ref{app.c}.

Since it is possible to scale $g$ and $\tilde{g}$ out of the large-$N$ IR theory by a simple rescaling of the fluctuation fields, $g\phi\to\phi$ and $\tilde{g}\varphi_{ij}\to\varphi_{ij}$,
we need to postulate that both Yukawa couplings are scale invariant, $\frac{dg}{d\ell} =\frac{d\tilde{g}}{d\ell} = 0$. From Eqs. (\ref{eq.dg}) and (\ref{eq.dgt}) and using our results for $z$ and $\Delta_\psi$
we obtain the scaling dimensions $\Delta_\phi = -4 +\eta_\phi$ and  $\Delta_\varphi = -4 +\eta_\varphi$ of the fluctuation fields where their resulting anomalous dimensions are
\begin{eqnarray}
\label{eq.etaphi}
\eta_\phi & = & 2\left(\Omega_g+\tilde{\Omega}_g-\Sigma_0-\tilde{\Sigma}_0   \right) \nn\\
& = &  -\frac{1}{3N}\frac{32(216\kappa-35)}{3\pi^2} \approx -\frac{6.077}{3N} \\
\label{eq.etavarphi}
\eta_\varphi & = & 2\left(\Omega_{\tilde{g}}+\tilde{\Omega}_{\tilde{g}}-\Sigma_0-\tilde{\Sigma}_0   \right) \nn\\
& = &  -\frac{1}{3N}\frac{8(240\kappa-41)}{\pi^2} \approx -\frac{3.353}{3N}.
\end{eqnarray}

To summarise, we have obtained the dynamical critical exponent $z$ (\ref{eq.z}), the fermion anomalous dimension $\eta_\psi$ (\ref{eq.etapsi}) and the anomalous dimensions $\eta_\phi$ (\ref{eq.etaphi}) and $\eta_\varphi$ (\ref{eq.etavarphi})
of the intra- and inter-valley fluctuation fields at one-loop order, which systematically accounts for contributions of order $1/(3N)$. At the topological phase transition the mass gap of the $3N$ Dirac fermions closes and we considered a GNY 
theory where both the dynamical intra- and inter-valley mass fluctuations are critical. 

Let us compare with the case where the inter-valley fluctuations are absent, $\tilde{g}=0$. This would simply mean that the inter-valley fluctuations are gapped and short-ranged. One would therefore expect to see a crossover from the universal 
behaviour with the set of critical exponents computed above to the universality of a GNY theory with $\tilde{g}=0$. We can obtain the critical exponents without inter-valley coupling by setting 
$\tilde{\Sigma}_0 = \tilde{\Sigma}_{x,y}=\Omega_{\tilde{g}}=\tilde{\Omega}_g=\tilde{\Omega}_{\tilde{g}}=0$, resulting in
\begin{equation}
z=1, \;\;\; \eta_\psi = \frac{1}{3N} \frac{4}{3\pi^2}, \;\;\; \eta_\phi = -\frac{1}{3N} \frac{32}{3\pi^2},
\end{equation}
which are the known critical exponents of the Ising-GNY theory in 2+1 dimensions in the limit of a large number of decoupled $3N$ copies of 2-component Dirac fermion fields \cite{Vasil'ev+1993, GRACEY+1991, GRACEY+1992, GRACEY+1994}.    

The comparison shows that the presence of critical or at least very soft inter-valley fluctuations has important consequences for the universal critical behaviour. Most importantly, it leads to a breaking of Lorentz 
invariance ($z>1$). While the changes of the fermion anomalous dimension are small, the anomalous dimensions of the bosonic fluctuation fields are significantly larger than that of the conventional GNY theory.

\section{Discussion}
\label{sec.conclude}

In conclusion we have addressed the nature of the field driven topological phase transition of the Kitaev QSL. Our mean-field results clarify the closing of the low energy band gap of the intermediate phase at the three $M$ points in the Brillouin zone, and the necessity to incorporate the hybridization with the high energy bands to account for the absence of fermion doubling in the IR, which is manifested through the presence of an odd number of Dirac cones. The hybridization makes the high energy bands topological, while permitting a redistribution of Berry curvature from the UV to the IR. 

We then performed a Wilson momentum shell renormalization group calculation in the GNY model to describe the nature of the quantum phase transition beyond mean-field. We showed that the inter-valley fluctuation channel among different $M$ points breaks Lorentz invariance and produces a dynamical exponent $z>1$. Inter-valley fluctuations produce small corrections to the mean-field critical exponents, with the exception of the anomalous dimension of the bosonic fields, where the effect is significant. The conclusion is that the intermediate topological phase transition of the Kitaev QSL at finite [111] field belongs  to a different universality class compared to the standard Ising GNY one.   

We acknowledge Arnaud Ralko for helpful discussions. BU acknowledges NSF grant DMR-2529526 for support.

%%%%%%%%%%%%%%%%%%%%%%%%%%%%%%
\begin{appendix}

\section{Low energy band Berry curvature}

 \begin{figure*}[t]
 \includegraphics[width=\linewidth]{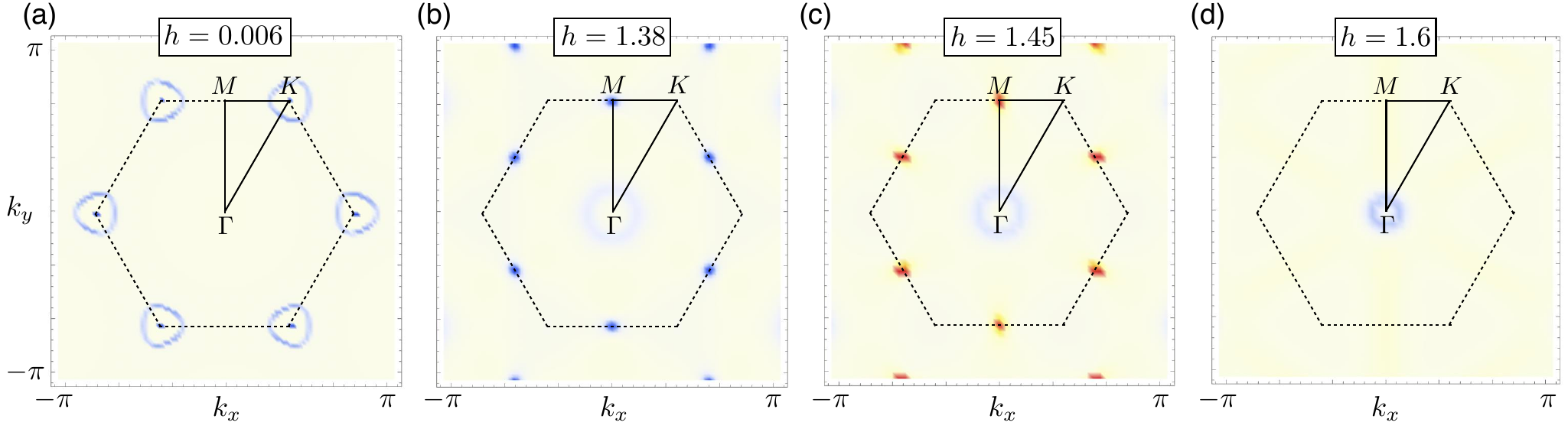}
 \caption{Berry curvature $\Omega(\bk)$ of the low-energy band for different values of magnetic field. (a) At small field ($h=0.006$) negative Berry curvature is concentrated at each $K$ point as well as around the 
 nodal line gap around $K$ from hybridisation with the flat bands.  (b) and (c) show the change of topological charge at the $M$ points from $-1/2$ to $1/2$ across the topological phase transition. The missing Berry 
 curvature of $-3/2$ is concentrated at a nodal line gap around $\Gamma$. (d) At $h=1.6$ Berry curvature is delocalised along the $\Gamma$-$M$ high-symmetry directions and the nodal-line gap around $\Gamma$ starts to contract.}
\label{figure6}
\end{figure*}

In  Fig.~\ref{figure6} we analyse the evolution of the  low-energy band Berry curvature $\Omega(\bk)$  for different field values.  At very small field, $h=0.006$ [see Fig.~\ref{figure6}(a)], we find negative spikes with  
topological charge of $-1/2$ at the $K$ points, as expected. In addition, there is a ring-like feature in the UV with $\Omega(\bk)<0$ around each $K$ point. This feature coincides with the hybridisation gap forming at the intersection line between the Dirac mode and 
the flat bands, see Fig.~\ref{figure2}. Each ring constitutes a topological charge of $-1$, resulting in a Chern number of ${\cal C}=2\times(-1/2)+2\times (-1)=-3$ of the low-energy band. 

Figs.~\ref{figure6}(b) and (c) show the Berry curvature $\Omega(\bk)$ at fields slightly below ($h=1.38$) and above ($h=1.45$) the critical value $h_c\approx 1.43$, where there is a topological phase transition. As  anticipated, the topological 
charge changes from $-1/2$ to $+1/2$ at each $M$ point, resulting in a Chern number change $\Delta \mathcal{C} =3$ across the transition. The missing Berry curvature of $-3/2$ is centred around a nodal line gap around $\Gamma$ between the 
low-energy and first exited bands. This nodal line gap remains intact across the transition and is clearly visible in the spectrum at $h_c$, shown in Fig.~\ref{figure2}(d).
Increasing the field to $h=1.6$ [see Fig.~\ref{figure6}(d)], the topological charge from the $M$ points seems to delocalise along the $\Gamma$-$M$ high-symmetry lines, along which the low-energy dispersion is practically flat
[see Fig.~\ref{figure2}(e)]. Moreover, the nodal line gap around $\Gamma$ seems to contract and move to lower energies.

Since the ground state remains topological at $h=1.6$, there needs to be at least one additional topological phase transition before the topologically trivial field polarised state is obtained at large $h$. 
Unfortunately, the numerics becomes unstable in this field range, which might be in part due to the confinement of Majorana fermions, which is not captured in the mean-field treatment.

\section{Bosonic self-energy corrections}
\label{app.a}

We will evaluate the regularised polarisation bubble diagram $f_{ij} (\vec{q})-f_{ij} (\vec{0})$ defined in Eq.~(\ref{eq.selfenergy0}). Inserting the expressions for the fermion Green function (\ref{eq.fermionGreen}), using that 
$\bm{\tau}_z^2=1$, $\bm{\tau}_z \bm{\tau}_x \bm{\tau}_z=-\bm{\tau}_x$, and $\bm{\tau}_z \bm{\tau}_y \bm{\tau}_z=-\bm{\tau}_y$ and taking the trace over pseudo-spin space, using that 
$\textrm{Tr} \left(\bm{\tau}_\alpha \bm{\tau}_\beta \right)=2\delta_{\alpha\beta}$, we obtain
\begin{equation}
f_{ij} (\vec{q}) = -2\int_{\vec{k}} \frac{F_{ij}(\vec{k},\vec{q})}{\vec{k}^2(\vec{k}+\vec{q})^2},
\end{equation}
where 
\begin{eqnarray}
F_{ij}(\vec{k},\vec{q}) & = &  k_0(k_0+q_0)+\cos\beta_{ij}\bk(\bk+\bq)\nn\\
& & -\sin\beta_{ij}(k_x q_y-k_y q_x).
\end{eqnarray}

Here $\beta_{ij}$ denotes the angle between the unit vectors $\hat\bn_i$ and $\hat\bn_j$. Subtracting 
\begin{equation}
f_{ij} (\vec{0}) = -2\int_{\vec{k}} \frac{k_0^2 +\cos\beta_{ij}\bk^2}{\vec{k}^4} = -\frac23\int_{\vec{k}} \frac{1 +2\cos\beta_{ij}}{\vec{k}^2}
\end{equation}
results in 
\begin{equation}
\tilde{f}_{ij} (\vec{q})=f_{ij} (\vec{q}) -f_{ij} (\vec{0}) = 2\int_{\vec{k}} \frac{\tilde{F}_{ij}(\vec{k},\vec{q})}{\vec{k}^2(\vec{k}+\vec{q})^2},
\end{equation}
with $\tilde{F}_{ij}(\vec{k},\vec{q})=\frac13 (1+2\cos\beta_{ij}) (\vec{k}+\vec{q})^2-F_{ij}(\vec{k},\vec{q})$. The IR behaviour of  $\tilde{f}_{ij} (\vec{q})$
is dominated by the small $\vec{k}$ contributions to the integral. We can therefore send the UV cut-off to infinity, what
enables us to use the standard Feynman parametrisation trick.  

We first introduce a dummy integration variable using the formula 
$1/(ab) = \int_0^1 dt/[t a+(1-t)b]^2$ with $a= (\vec{k}+\vec{q})^2$ and $b=\vec{k}^2$, followed by a shift of the frequency-momentum vector, $\vec{p}=\vec{k}+t\vec{q}$, to obtain 
\begin{equation}
\tilde{f}_{ij} (\vec{q})= 2\int_0^1 dt \int_{\vec{p}} \frac{\tilde{F}_{ij}(\vec{p}-t\vec{q},\vec{q})}{[\vec{p}^2+t(1-t)\vec{q}^2]^2}.
\end{equation}
The denominator is now rotationally symmetric in $\vec{p}$ and, as a result, the terms in $\tilde{F}_{ij}(\vec{p}-t\vec{q},\vec{q})$ that are linear in $\vec{p}$ vanish under integration. Moreover, the terms
that are quadratic in the components of $\vec{p}$ cancel each other under integration, 
leaving the remaining integral 
\begin{eqnarray}
\tilde{f}_{ij} (\vec{q}) & = &  2\int_0^1 dt  \Big\{ (1-t)^2 \frac{1+2\cos\beta_{ij}}{3} \vec{q}^2   \\
& & + t(1-t) (q_0^2+\cos\beta_{ij}\bq^2) \Big\}  \int_{\vec{p}} \frac{1}{[\vec{p}^2+t(1-t)\vec{q}^2]^2}.\nn
\end{eqnarray}
After carrying out the radially symmetric, three dimensional $\vec{p}$ integral we obtain
\begin{eqnarray}
\tilde{f}_{ij} (\vec{q}) & = &  \frac{1}{4\pi|\vec{q}|}\int_0^1 \frac{dt}{\sqrt{t(1-t)}} \Big\{ (1-t)^2 \frac{1+2\cos\beta_{ij}}{3} \vec{q}^2   \nn\\
& & + t(1-t) (q_0^2+\cos\beta_{ij}\bq^2) \Big\}.
\end{eqnarray}

\noindent
The integrals over $t$ are elementary, 
\begin{equation}
\int_0^1 dt  \frac{(1-t)^2}{\sqrt{t(1-t)}} =\frac{3\pi}{8}\;\;\textrm{and}\;\; \int_0^1 dt  \frac{t(1-t)}{\sqrt{t(1-t)}} =\frac{\pi}{8},
\end{equation}
resulting in Eq.~(\ref{eq.selfenergy}).

\section{Fermion self energy corrections}
\label{app.b}

Let us first evaluate the fermion self energy correction $d\bm{\Sigma}_i(\vec{k})$ (\ref{eq.fermion_ren1}) from the intra-valley fluctuations. After Taylor expansion in external frequencies and 
momenta to linear order, the coefficients $\Sigma_n d\ell$ in Eq.~(\ref{eq.fermion_ren3}) are obtained as the following shell integrals, 
\begin{eqnarray}
\Sigma_0 d\ell & = & \frac{g^2}{3N} \int_{\vec{q}}^> \frac{D(\vec{q})}{\vec{q}^2} \left(1-2\frac{q_0^2}{\vec{q}^2} \right)\\
\Sigma_x d\ell & = & \frac{g^2}{3N} \int_{\vec{q}}^> \frac{D(\vec{q})}{\vec{q}^2} \left(1-2\frac{[(\hat{\bn}_i\times \ez)\cdot\bq]^2}{\vec{q}^2} \right)\\
\Sigma_y d\ell & = & \frac{g^2}{3N} \int_{\vec{q}}^> \frac{D(\vec{q})}{\vec{q}^2} \left(1-2\frac{(\hat{\bn}_i\cdot\bq)^2}{\vec{q}^2} \right).
\end{eqnarray}

Since $D(\vec{q})$ is rotationally symmetric it follows that $\Sigma_0=\Sigma_x=\Sigma_y$ and
\begin{eqnarray}
\Sigma_n d\ell  & = &   \frac{g^2}{3N} \int_{\vec{q}}^> \frac{D(\vec{q})}{\vec{q}^2} \times \left(1-\frac23\right) = \frac{1}{3N} \frac{8}{3} \int_{\vec{q}}^> \frac{1}{|\vec{q}|^3} \nn\\ 
& = &  \frac{1}{3N} \frac{4}{3\pi^2} \int_{\Lambda e^{-d\ell}}^{\Lambda} \frac{dq}{q} =  \frac{1}{3N} \frac{4}{3\pi^2} d\ell.
\end{eqnarray}

In the case of the inter-valley contributions an additional complication arises from the fact that  $d\tilde{\bm{\Sigma}}_i(\vec{k})$  (\ref{eq.fermion_ren2}) involves a sum over 
Green functions $\bG_j$ from neighbouring valleys. However, using that $\sum_{j(\neq i)} \hat{\bn}_j = \hat{\bn}_i$ we obtain
\begin{eqnarray}
& & -\sum_{j(\neq i)}\bm{\tau}_z \bG_j(\vec{k}+\vec{q})\bm{\tau}_z\\
& & = \frac{-2i(k_0+q_0) +(\hat{\bn}_i\times \ez)\cdot(\bk+\bq)\bm{\tau}_x + \hat{\bn}_i\cdot(\bk+\bq)\bm{\tau}_y}{(\vec{k}+\vec{q})^2}.\nn
\end{eqnarray}

Proceeding with the Taylor expansion as in the intra-valley case, this results in 
\begin{eqnarray}
\tilde{\Sigma}_0 d\ell & = & \frac{\tilde{g}^2}{3N} \int_{\vec{q}}^> \frac{\tilde{D}(\vec{q})}{\vec{q}^2} \left(1-2\frac{q_0^2}{\vec{q}^2} \right)\\
\tilde{\Sigma}_x d\ell & = & \frac{\tilde{g}^2}{6N} \int_{\vec{q}}^> \frac{\tilde{D}(\vec{q})}{\vec{q}^2} \left(1-2\frac{[(\hat{\bn}_i\times \ez)\cdot\bq]^2}{\vec{q}^2} \right)\\
\tilde{\Sigma}_y d\ell & = & \frac{\tilde{g}^2}{6N} \int_{\vec{q}}^> \frac{\tilde{D}(\vec{q})}{\vec{q}^2} \left(1-2\frac{(\hat{\bn}_i\cdot\bq)^2}{\vec{q}^2} \right).
\end{eqnarray}

\noindent
Inserting $\tilde{D}(\vec{q})$ and using spherical coordinates we obtain 
\begin{eqnarray}
\tilde{\Sigma}_0 d\ell & = & \frac{32}{3N} \int_{\vec{q}}^> \frac{1}{|\vec{q}]^3}\frac{1-2 \frac{q_0^2}{\vec{q}^2}}{1-\frac16 \frac{\bq^2}{\vec{q}^2}}\nn\\
& = & \frac{1}{3N} \frac{8}{\pi^2} d\ell \int_0^\pi d\theta \sin\theta \frac{1-2\cos^2\theta}{1-\frac16 \sin^2\theta}\nn\\
& = & \frac{1}{3N}\frac{96}{\pi^2}\left[\frac{11}{\sqrt{5}} \arctan\left(\frac{1}{\sqrt{5}} \right) -2  \right]d\ell\\
& = & \frac{1}{3N}\frac{96(11\kappa-2)}{\pi^2}d\ell\,
\end{eqnarray}
for the frequency coefficient, where we have defined 
\begin{equation}
\kappa = \frac{1}{\sqrt{5}}\arctan\left( \frac{1}{\sqrt{5}}  \right).
\end{equation}
Likewise, for the coefficients of the spatial momentum terms we obtain
\begin{eqnarray}
\tilde{\Sigma}_{x,y} d\ell & = & \frac{32}{6N} \int_{\vec{q}}^> \frac{1}{|\vec{q}]^3}\frac{1-\frac{\bq^2}{\vec{q}^2}}{1-\frac16 \frac{\bq^2}{\vec{q}^2}}\nn\\
& = & \frac{1}{3N} \frac{4}{\pi^2} d\ell \int_0^\pi d\theta \sin\theta \frac{1-\sin^2\theta}{1-\frac16 \sin^2\theta}\nn\\
& = & \frac{1}{3N}\frac{48(1-5\kappa)}{\pi^2}d\ell.
\end{eqnarray}

\section{Corrections to Yukawa couplings}
\label{app.c}

The loop corrections to the Yukawa couplings can be written as $\Omega_g d\ell = \frac{1}{3N}g_{ii} d\ell$, $\tilde{\Omega}_g d\ell = \frac{1}{3N}\tilde{g}_{ii} d\ell$, $\Omega_{\tilde{g}} d\ell = \frac{1}{3N}g_{i\neq j} d\ell$, and 
$\tilde{\Omega}_{\tilde{g}} d\ell = \frac{1}{6N} \tilde{g}_{i\neq j} d\ell$, where
\begin{eqnarray}
g_{ij} d\ell & = & g^2 \int_{\vec{q}}^> D(\vec{q}) \bG_i(\vec{q}) \bm{\tau}_z \bG_j(\vec{q}) \bm{\tau}_z\\
\tilde{g}_{ij} d\ell & = & \tilde{g}^2 \int_{\vec{q}}^> \tilde{D}(\vec{q}) \bG_i(\vec{q}) \bm{\tau}_z \bG_j(\vec{q}) \bm{\tau}_z.
\end{eqnarray}
Using that 
\begin{eqnarray}
\bG_i(\vec{q}) \bm{\tau}_z \bG_j(\vec{q}) \bm{\tau}_z & = &  - \frac{1}{\vec{q}^2}\left[1-(1-\cos\beta_{ij})\frac{\bq^2}{\vec{q}^2}   \right]   \\
& & +\textrm{terms that vanish under int.},\nn
\end{eqnarray}
where $\cos\beta_{ij}=\hat{\bn}_i\cdot \hat{\bn}_j$, we can use the spherical symmetry of $D(\vec{q})$ to compute $g_{ij}d\ell$,
\begin{eqnarray}
g_{ij}d\ell  & = &  -8 \int_{\vec{q}}^> \frac{1}{|\vec{q}|^3}\left[1-\frac23 (1-\cos\beta_{ij})    \right] \nn\\
& = & -\frac{4}{\pi}^2 \left[1-\frac23 (1-\cos\beta_{ij})    \right] d\ell.
\end{eqnarray}

Using that $\cos\beta_{ii}=1$ and $\cos\beta_{i\neq j}=1/2$, this results in the expressions (\ref{eq.Yuk1}) and (\ref{eq.Yuk3}) for $\Omega_g d\ell$ and $\Omega_{\tilde{g}} d\ell$. 
For the integrals involving $\tilde{D}(\vec{q})$ we obtain
\begin{eqnarray}
\tilde{g}_{ij} d\ell & = & -32 \int_{\vec{q}}^> \frac{1}{|\vec{q}|^3} \frac{1-(1-\cos\beta_{ij})\frac{\bq^2}{\vec{q}^2}}{1-\frac16 \frac{\bq^2}{\vec{q}^2}}\nn\\
& = & -\frac{8}{\pi^2} d\ell \int_0^\pi d\theta \sin\theta \frac{1-(1-\cos\beta_{ij})\sin^2\theta}{1-\frac16 \sin^2\theta}\nn\\
& = & -\frac{8}{\pi^2} d\ell \left\{ \begin{array}{c} 12\kappa \quad (i=j) \\ 6(1-4\kappa)\quad (i\neq j)     \end{array}   \right.,
\end{eqnarray}
which reproduces  Eqs. (\ref{eq.Yuk2}) and (\ref{eq.Yuk4}) for $\tilde{\Omega}_g d\ell$ and $\tilde{\Omega}_{\tilde{g}} d\ell$. 

\end{appendix}

\end{document}